# Violence in Guatemala pushes adults and children to seek work in Mexico

Roxana Gutiérrez-Romero[1]

**Abstract**

This article estimates the impact of violence on emigration crossings from Guatemala to Mexico as final destination during 2009-2017. To identify causal effects, we use as instruments the variation in deforestation in Guatemala, and the seizing of cocaine in Colombia. We argue that criminal organizations deforest land in Guatemala, fueling violence and leading to emigration, particularly during exogenous supply shocks to cocaine. A one-point increase in the homicide rate differential between Guatemalan municipalities and Mexico, leads to 211 additional emigration crossings made by male adults. This rise in violence, also leads to 20 extra emigration crossings made by children.

**Wordcount**: 8,000

[1] Professor of Policy and Quantitative Methods at Queen Mary University of London. Mile End Campus, Bancroft Building, 4th Floor, London, UK, E1 4NS. r.gutierrez@qmul.ac.uk. I acknowledge funding from the Global Challenges Research Fund (GCRF) [RE-CL-2021-01]. I am also grateful to Daniel Acosta Chávez, Tania Rodríguez, Yunuen Rodríguez, Nayeli Salgado and David Aban Tamayo, for excellent research assistance.

# 1. Introduction

Most international migration takes place from developing countries to advanced economies. Given that emigrating to another country is costly, only a small percentage of the population can make the move. But, for some very poor households in the Global South that lack funds, migrating to a nearby developing country might still represent a viable option. Moving to a country with similar education system and labor market can be particularly attractive to children and young adults often exposed to high risks of deprivation, violence and environmental disasters. South-South migration accounts today for about 40% of all outward international migration and is an increasingly important source of development (Ratha and Shaw 2007). However, the leading factors driving these new population movements are still debated (Campillo-Carrete 2013). In under-researched regions such as Central America, it is even more important to understand the relevant factors driving emigration to develop an effective migration strategy in both sender and host communities. Over the last 30 years, the number of migrants from Central America has increased from 6.8 to nearly 16.2 million (UNDESA 2020). Guatemala experienced the second highest rise in emigration in the region, with an increase of nearly 295%, after Honduras (with 530%). The literature has thus far focused on analyzing these emigration flows to the United States, overlooking that Mexico also became an important final destination.

This article fills an important gap in the literature by evaluating the causal relationship between rises in local violence on adult and children emigrating from Guatemala to Mexico. We examine the Survey of Migration to the Southern Border of Mexico, known as EMIF Sur, during 2009-2017. This survey, the largest survey on transit migration in the region, provides representative estimates of the number of emigration crossings, documented or not, made by land by people seeking jobs or moving to the USA or Mexico for a period of a month or longer (COLEF 2013).[2] Our analysis, unlike the majority of recent studies does not analyze child migrants who were apprehended and deported at the USA border. While it is crucial to analyze the growing phenomenon of unaccompanied child migrants leaving, being apprehended and deported at the USA border, it is also important to broaden the analysis to include adults and child migrants crossing the Guatemala-Mexican border. By considering adult and child migrants who travel to Mexico as final destination, help us provide a more comprehensive picture of this migration crisis and avoid underestimating its severity.

---

[2] These emigration flows refer to the number of crossings made and not the number of people migrating. This is an important distinction as people at the border migrate several times a year for seasonal agricultural work. Still, we can provide a more comprehensive picture of the net impact of violence on population movements.

We focus exclusively on survey respondents that lived in Guatemala before migrating, which comprises about 95% of the EMIF Sur sample. We also restrict our analysis to the period of 2009-2017 for which we have annual data on homicide rates at municipality level, as well as relevant statistics on poverty levels. Migrants leaving during this period would not qualify to gain work permit visas given in the Obama's administration during the Deferred Action for Childhood Arrivals (DACA) program. The overwhelming majority of these migrants also started their journey before the Trump administration launched a pilot program in Arizona to separate families at the USA border in May 2017. Thus, the period analyzed allows us to understand to what extent violence and socio-economic factors in Guatemala drove the unprecedented flows of emigrants.

A key challenge in the child migration literature is the likely endogenous relationship between emigration decisions and local violence. Violence can respond to emigration flows (Ambrosius 2021). According to recent estimates roughly half of Central American migrants leave their children behind (Abuelafia, Del Carmen, and Ruiz-Arranz 2019). These children are at risk of being recruited by local gangs which risks fueling crime and violence at home. It is also possible that the growing trend of adults and children migrating north could reduce violence, with fewer potential victims. An important contribution of this article is to unravel the effect of violence on emigration by exploiting the exogenous variation in the seizing of cocaine in Colombia and the variation of local deforestation in Guatemala.

Whenever the Colombian government seizes cocaine it disrupts the supply chain of illicit drugs. The reduction in supply leads to rises in the retail prices of cocaine, and fuels conflicts in the region as criminal organizations fight over a more lucrative market (Castillo, Mejía, and Restrepo 2020; Sarrica 2008). Cocaine gets trafficked through Central America, the corridor that transports 90% of the cocaine consumed worldwide (UNODC 2010). Drug trafficking organizations legitimize some of their illicit profits by deforesting land and converting it into cattle ranching, agro-industrial plantations, or other infrastructure such as roads and clandestine airstrips to assist their trafficking (Hodgdon et al. 2015; McSweeney et al. 2014). Drug trafficking organizations in Guatemala operate in roughly half of the country's territory and are responsible for at least 15-30% of the deforestation that Central America has experienced since the mid-2000s (Nellemann 2012; Sesnie et al. 2017). Our logic is that local deforestation triggers more violence in Guatemala, relatively Mexico, indirectly leading to more emigration flows. We show that this violence is increased particularly when the supply of cocaine is affected by seizures of cocaine production in Colombia.

Contrary to the dominant narrative in news reports and literature, we find that about 1% of the EMIF Sur sample are on their transit to the USA. The rest are border workers migrating to Mexico as their intended final destination, and often crossing the border several times a year. At the very least 60% of these crossings are undocumented. We show there is evidence of endogeneity. But using instrumental variables we demonstrate that a one-point increase in the local homicide rates, relative to the average level in Mexico, increases the number of emigration crossings made by male adults, children, and unaccompanied minors to Mexico (by 211, 20 and 13 respectively).

## 2. Drivers of South-South migration

*2.1 Push and Pull Factors of South-South Migration*

Over the last two decades Mexico has become an important recipient and final intended destination for many migrants in the region, particularly from Central America (COLEF 2013). Mexico's close proximity (geographical, cultural and historical) makes it an attractive destination and a relatively safe refuge for many Central American migrants. But the reasons why both adult and child migration have substantially increased in the region are still debated in the literature (Hoekstra and Orozco-Aleman 2021; Clemens 2021; Amuedo-Dorantes, Pozo, and Puttitanun 2015). The extensive migration literature agrees that civil war and other large scale-conflicts are major push factors for emigration and forced displacement in several developing regions (Chetail 2014). For the Central American case, critics, point out that violence in the region has been endemically high for decades, and that the recent migration crises in the region can instead be explained by recent changes in USA immigration policy (Herridge 2015; Swanson and Torres 2016).

Some argue that human smugglers have misled families to send their children to the USA to get work visas under then-President Obama's Deferred Action for Childhood Arrivals (DACA) programme (EPIC 2014). This large-scale immigration policy granted people with unlawful presence in the USA a renewable two-year period of deferred action from deportation and a work permit. There were several eligibility requirements such as having lived continuously in the USA since June 15 2007, being under age 31 on as of June 15, 2012, and having an unlawful presence in the USA after entering the country before their 16th birthday. Nonetheless, there is no strong evidence to suggest this program affected the migration decisions of Central Americans because the growing trend of migrant children pre-dates the announcement of DACA (American Immigration Council 2014). Moreover, none of the recent arrivals would qualify to be granted a visa.

Even if families were misinformed, the large flows of deportees being sent back to Central America would help to update and re-align expectations (CBP 2017). Since 2014, Mexico also strengthened substantially its border security with Guatemala leading to more deportations of undocumented migrants, particularly during Trump's administration (Fredrick 2019).

*2.2 Limitations from Current Emigration Literature*

More formally, a large literature has analyzed the impact of USA border enforcement on irregular emigration flows in Central America (e.g. Amuedo-Dorantes et al. 2015; Espenshade and Acevedo 1995; Hagan et al. 2008). Some studies find that apprehension and deportations from the USA deter undocumented emigration flows from Central America (Martínez Flores 2020). Others, find that Central American parents that get forcedly separated from their children at the USA border are more likely to intend to return to the USA, presumably undocumented (Amuedo-Dorantes et al., 2015). The emphasis of the literature in analyzing the effects of USA policy on child migrants ignores two key aspects: the role of Mexico as another important destination, and how violence affects both children and adults, as discussed next.

Mexico has become an important destination for many Central American migrants, particularly border workers. According to a recent survey of over 4000 unaccompanied minors in USA immigration custody, organized criminal violence and domestic violence are the two main leading causes of child migration in Mexico, Honduras, Guatemala and El Salvador (UNHCR 2014). In Guatemala, close to 50% children interviewed report having to cope with violence in both their home and community in addition to severe socio-economic deprivation. These self-reported views coincide with other migration studies in the region and similar developing countries. Child neglect at the hands of parents has been found to motivate children to migrate unaccompanied in India, for instance (Iversen 2002). The loss of one or both parents has been found to trigger child migration in other developing countries such as Bangladesh (Heissler 2012). Children are at a very high risk of victimization and death in Latin America, but even more acute in Central America (Wong 2014). The region also has the highest rate of youth homicide. Given the high mortality of young adults, children are more at risk of becoming orphans at a younger age in the region than their counterparts in the Global North.

Another limitation of the growing migration literature is the focus of the role of violence on child migration, without exploring in depth whether violence equally impacts adults. Young adults might have even more funds to afford longer journeys to safer destinations. Although violence in Mexico has increased substantially since the mid-2000s, violence has remained at lower levels than in much of Central America, and with an even higher income per capita, Mexico has become an attractive destination. In addition to fleeing violence, young adults and children might migrate temporarily or more permanently for economic purposes.

Most poor people in the Global South live in rural areas where they are reliant on seasonal work, where migration within the country might be common. Landless migrants spend a considerable time migrating for short periods to nearby cities and neighboring developing countries for seasonal rural work. Thus, children in developing countries can spend prolonged periods of time without one or both of their parents. For instance, the percentage of children living in migrating households is as high as 60% in Tanzania and 80% in Mali (Whitehead, Hashim, and Iversen 2007). With these recurrent migration patterns children can perceive migration as an important way of economic survival and enhance their level of agency in deciding to migrate for family reunification or on their own. Natural disasters can also lead to profound economic losses and potentially lead to emigration (Meyer 2021).

Based on this discussion we will test whether rises in local violence have increased adult and child migration from Central America. We will focus on Guatemalans, who form the great majority of those described in the large migration survey analyzed here. Since Mexico is also an increasingly violent country, we will test whether rises in the local homicide rate in Guatemalan municipalities, relative to the average in Mexico, are relevant as a push factor for emigration. As explained below, our period of analysis of 2009-2017 is outside the scope of the DACA immigration policy, and family-child separations implemented during the Trump administration. Thus, our period is ideal to test the role of violence in driving migration decisions.

### 3. Violence in Guatemala

The Guatemala-Mexican migration corridor is an important case study for understanding how poverty and violence affect population movements, particularly how young people in the Global South cope with the legacies of civil war. Guatemala and Mexico share a substantial history through the Mayan civilization, the conquest, and the post-Spanish colonization era. After Mexico gained independence from the Spanish Empire in 1821, Central American states came under Mexican administration until 1823, thereafter gaining their independence. The modern-day border shared between Guatemala and Mexico was agreed in 1882, being 871 kilometers long and including long stretches of the Usumacinta, Salinas, and the Suchiate rivers (Figure 1).

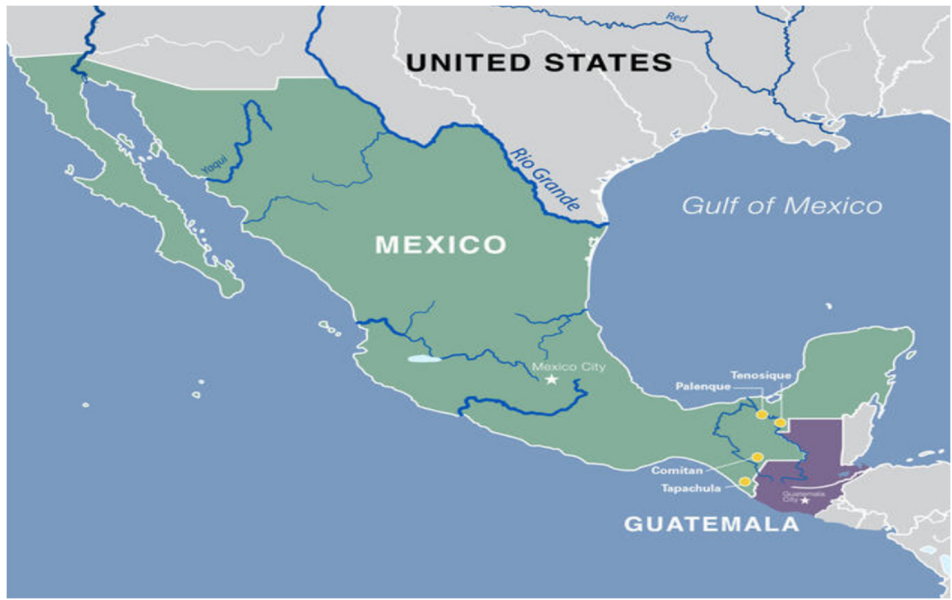

Fig 1. Mexico-Guatemala border.

Violence in Guatemala has been a persistent issue even after the end of 36-year civil war, over two decades ago. The end of civil war has not reversed the persistently high levels of violence and poverty in the country. Guatemala remains one of the poorest countries in Latin America with about 50% living on less than $5 dollars a day (Atamanov et al. 2018). The legacies of the civil war have also left Guatemala in the top-ten most violent nations in the world, with a homicide rate of 33.6 per 100,000 inhabitants in 2017 (Roser and Ritchie 2019).

Although violence in Central America is not new, homicide rates rose in the 2000s in Central America as the region became the main transit corridor for South American narcotics to the USA (Hodgdon et al. 2015). Drug trafficking organizations have since engaged in a turf war with one another to secure trafficking routes. Organized crime and local gangs, such as the *Mara Salvatrucha* and the *M-18* in Central America, continue to recruit young adults to further their illicit activities of extortion and local drug distribution (Swanson and Torres 2016). While violence in Guatemala is high, its southern neighbors suffer from even higher levels. Mexico has also suffered from high levels of violence and drug trafficking, but the homicide rate in Guatemala have remained higher (Figure 2). Thus, for Guatemalans struggling to cope with poverty, violence, and lack of opportunities, the only route out is heading north, to Mexico or the USA.

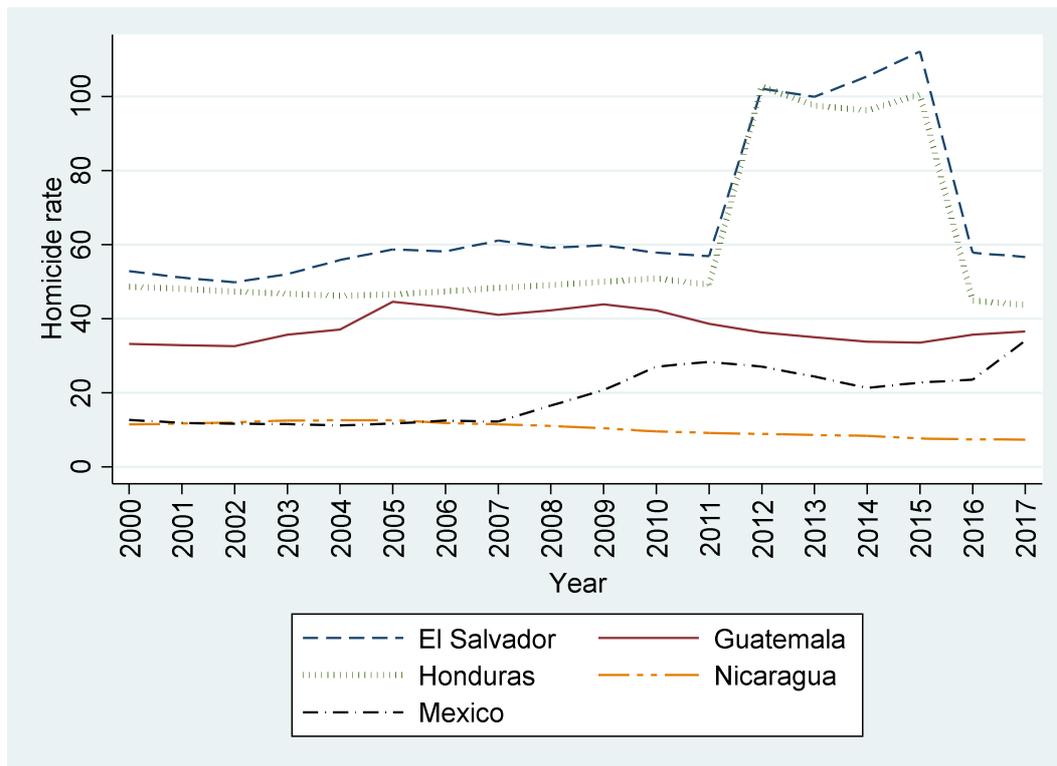

Fig 2. Homicides in El Salvador, Guatemala, Honduras, Mexico and Nicaragua during 2000-2017.

Source: Our world in data

## 4. Data and Descriptive Statistics

*4.1 Data Sources*

We rely on the Survey of Migration to the Southern Border of Mexico, known as EMIF Sur, conducted by *El Colegio de la Frontera* (COLEF) in collaboration with various government institutions. The EMIF Sur is the most comprehensive and representative survey of transit migration from Central America. The survey provides a nationally representative number of the total crossings by land, whether documented or not, made by people aged 14 and over that are migrating to Mexico or the USA for work, living purposes or visiting family for a period of over a month. Note that the estimates provided refer to the number of crossings and not the number of people migrating. This is an important distinction to bear in mind, since people might cross the border several times a year for work purposes, particularly those living near the border. The majority of crossings (60%) are made by undocumented migrants, and the rest were made by people who either had valid visas or were travelling to arrange such documentation.

The EMIF Sur uses a sampling framework suitable for mobile populations (COLEF 2013).[3] Migrants are interviewed in the most important crossings at the Mexican-Guatemala border, including custom inspections points and bus stations most commonly used for undocumented migrants. The EMIF Sur includes information only on people that are about to migrate. Thus, it is not possible to compare migrants to non-migrants. Nonetheless, we analyze the impact of violence on emigration crossings at municipality level in Guatemala. We focus on the crossings made by people that at the time of the survey were residing in Guatemala, which represents about 95% of the sample. We analyze residents of Guatemala because only for this population it is possible to ascertain in the EMIF Sur the municipality and department where they were living. Thus, we can consider the level of violence and poverty of their areas of origin.

Table 1. Main reason Guatemalans cross Mexican border during 2009-2017

| Main reason | All | | | Male | | | Female | | | Children | | |
|---|---|---|---|---|---|---|---|---|---|---|---|---|
| | Freq. | Sampling weights Freq. | Percent | Freq. | Sampling weights Freq. | Percent | Freq. | Sampling weights Freq. | Percent | Freq. | Sampling weights Freq. | Percent |
| To work in Mexico | 76,496 | 5,565,012 | 99.24 | 65,021 | 4,455,548 | 99.31 | 8,440 | 887,457 | 98.91 | 3,035 | 222,007 | 99.152 |
| To live in Mexico | 279 | 14,430 | 0.26 | 226 | 10,707 | 0.24 | 42 | 3,094 | 0.34 | 11 | 629 | 0.281 |
| To visit family or friends in Mexico | 112 | 8,067 | 0.14 | 69 | 4,739 | 0.11 | 29 | 2,691 | 0.30 | 14 | 637 | 0.284 |
| To know Mexico | 11 | 1,432 | 0.03 | 5 | 514 | 0.01 | 5 | 896 | 0.10 | 1 | 22 | 0.010 |
| To work in USA | 111 | 18,290 | 0.33 | 99 | 14,549 | 0.32 | 8 | 3,130 | 0.35 | 4 | 611 | 0.273 |
| To visit family or friends in USA | 2 | 331 | 0.01 | 2 | 331 | 0.01 | 0 | 0 | 0.00 | 0 | 0 | 0.000 |
| To know USA | 1 | 59 | 0.00 | 1 | 59 | 0.00 | 0 | 0 | 0.00 | 0 | 0 | 0.000 |
| Total | 77,012 | 5,607,621 | 100.00 | 65,423 | 4,486,447 | 100.00 | 8,524 | 897,268 | 100.00 | 3,065 | 223,906 | 100.00 |

Source: EMIF Sur.

We restrict our analysis to the period of 2009-2017 for which we have annual data on homicide rates at municipality level provided by the Guatemalan Police. During this period, violence rose substantially in both Mexico and Central America because of the ongoing drug cartels' turf war. Migrants therefore need to weigh the risks of moving from one violent country to another one also experiencing growing violence, and where migrants are known to be victims of extorsion and kidnapping (Swanson and Torres 2016).

---

[3] For further information, including how the COLEF obtains the population sampling weights used to estimate the total number of crossings made by the migrant population, see https://www.colef.mx/emif/bases.html.

Table 1 shows the number of crossings made by people residing in Guatemala to the Mexican or the USA border. During our period of analysis, 2009-2017, the EMIF Sur sample recorded a total of 77,012 crossings made by interviewed people. Only a minority of these crossings, less than 1%, were made by people that intended to migrate to the USA. Instead, most of the crossings were made by people that migrated to Mexico as their final destination and for work purpose. The EMIF Sur provides sampling weights to yield an overall estimate of the total number of crossings made by land, whether documented or not, that their sample represents. Using these sampling weights, it is estimated that a total of 5,607,621 crossings were made during 2009-2017. Most of these crossings (5,565,012) were made from Guatemala to Mexico as a final destination. It is worth noting that these are not the number of people migrating but the number of estimated emigration crossings in that period analyzed.

Table 2. Summary statistics of EMIF Sur

| Number of emigration crossings during 2009-2017 made by: | Unweighted observations | Weighted observations |
| --- | --- | --- |
| **Migrating from Guatemala to Mexico** | | |
| Adults | 73,837 | 5,365,646 |
| Men | 65,321 | 4,471,508 |
| Women | 8,516 | 894,138 |
| Children | 3,061 | 223,295 |
| Girls | 278 | 27,831 |
| Unaccompanied children | 2,203 | 162,668 |
| **Migrating from Guatemala to Mexico or the USA** | | |
| Unaccompanied children | 2,220 | 164,869 |
| Unaccompanied girls | 220 | 22,428 |

Source: EMIF Sur.

Table 2 shows the unweighted and weighted number of crossings made by adults, male, women, children and unaccompanied children from Guatemala to Mexico during 2009-2017. Nearly 90% of emigration crossings made by adults were made by men. According to the EMIF Sur, these adults were on average age 33, married (70%), heads of their household (70%), with low level of education attainment (31% without schooling and 60% with only primary). Only a minority of these emigrants had a job before departing (26%).

The EMIF Sur also registered 3,061 crossings made by minors (aged between 14 and 17) intended to Mexico as the final destination. Using the EMIF Sur sampling weights, these crossings represent about 223,295 crossings in total. Most of these crossings (90%) were made by boys, and (70%) by unaccompanied minors. Table 2 also shows that if we consider instead all the crossings made by unaccompanied minors to Mexico or the USA, most (99%) were made to Mexico as the intended final destination and by boys.

*4.2 Descriptive Evidence on Number of Emigrating Crossings*

Figure 3 shows the annual number of crossings made by Guatemalan residents to Mexico and the USA as their final destination. The number of crossings have already been scaled up using the EMIF sampling weights. The panel on the left of Figure 3 shows a sharp increase in the number of crossings made by Guatemalan residents migrating into Mexico during 2009-2014, followed by a gradual decline since 2015, and then a steep decline of 67% in 2017. The panel on the right in Figure 3 shows that the number of crossings from Guatemalan residents to the USA also fell sharply in 2016 and 2017 compared to previous years. The reduction in these emigration crossings during 2016 and 2017 does not appear to be driven by an economic crisis in Guatemala.

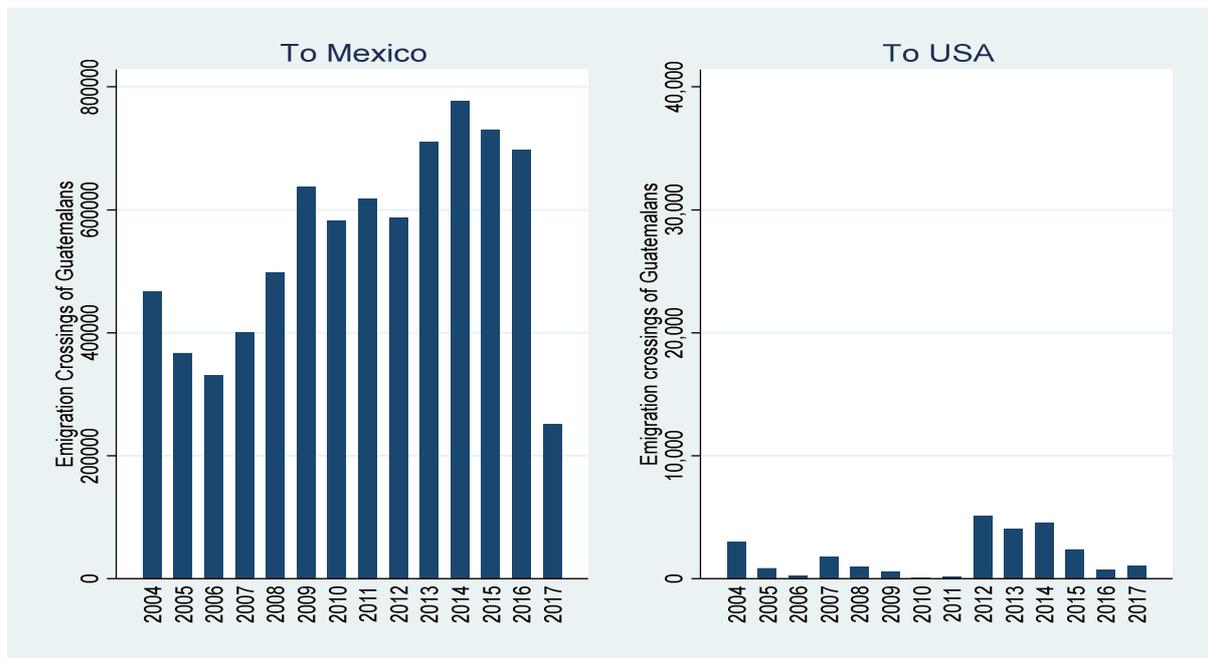

Fig 3. Number of crossings made by Guatemalans to Mexico and to USA during 2009-2017. Source: Author's own estimates. EMIF Sur, weighted data.

Guatemala's Gross National Income (GNI) per capita has had a modest growth during our period of analysis, albeit still at a much lower level than that those in Mexico and USA (Figure 4). Instead, as Hoekstra & Orozco-Aleman (2021) argue, the reduction in emigration flows observed during 2017 seems to be related to the unexpected election of Donald Trump in the USA. These authors identify a similar reduction in the emigration flows from Central Americans using the EMIF North survey, which monitors the crossings over Mexico's northern border. They compare the migration flows before and after 2016 and conclude that the unexpected election of Trump, who took a firm stance against undocumented migration, temporarily reduced emigration flows from Central America.

As mentioned earlier, most people interviewed by the EMIF Sur claim that are crossing to the Mexico and not to the USA. However, the deportation patterns from Mexico and USA suggest that some of these Guatemalan emigrants will eventually cross or attempt to cross into the USA (CBP 2017). Thus, the rise in deportation from the USA since 2017 can also explain why the crossings to Mexico also suffered a sharp reduction that year. Moreover, since 2017, Mexico has increased its border enforcement on its south border, in close collaboration with the USA (David 2018).

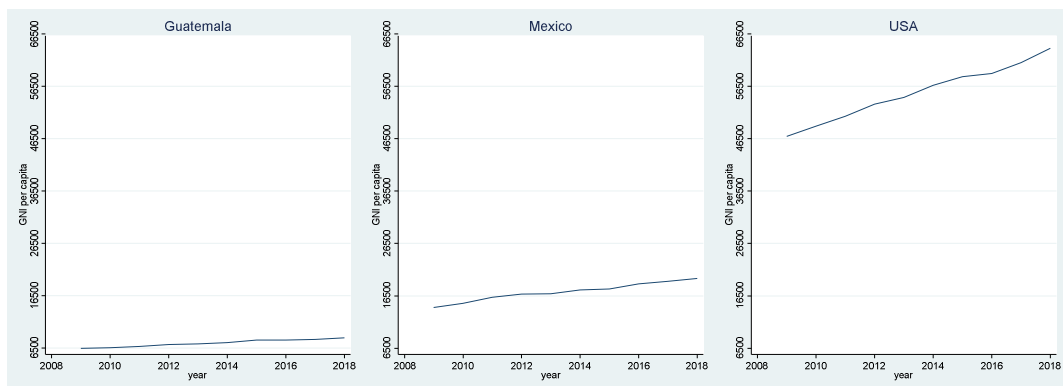

Fig 4. Gross National Income (GNI) per capita at purchasing power parity in Guatemala, Mexico and USA during 2009-2017.
Source: World Bank

To start unravelling the role of local violence and poverty as potential push factors for emigration, we show next the geographic distribution of homicide rates and emigration crossings made by adults, women and children at municipality of origin in Guatemala during 2009-2017. Figure 5 shows that emigration flows were primarily coming from municipalities that are closest to the Mexican border, which also have the highest levels of poverty and the lowest levels of violence. Children leaving Guatemala without parents or anyone else originated from areas located along the border with Mexico. Nonetheless, there is a relatively high concentration of child migrants in a few municipalities in the center and east, towards the Honduras border. These areas have the highest rates of homicides in the country, and moderately high levels of poverty.

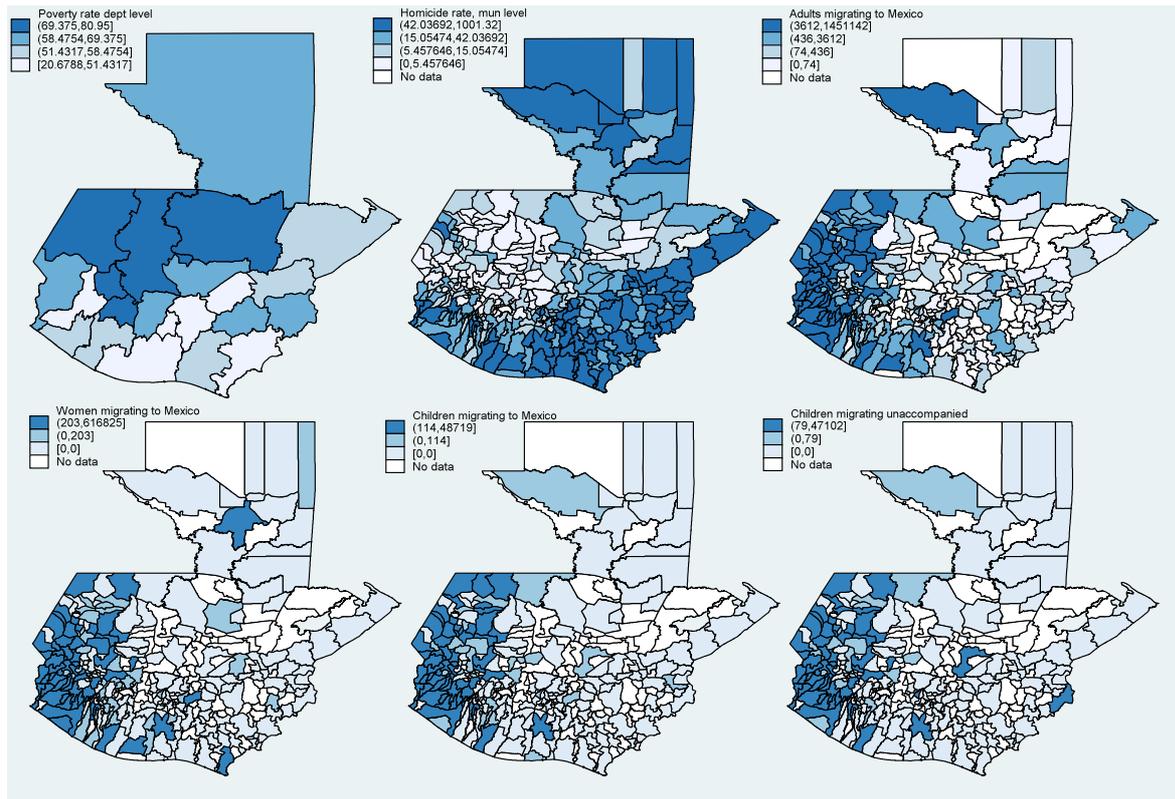

Fig 5. Homicide rates, number of emigration crossings made by Guatemalans at municipality level, and poverty rate by departments in Guatemala during 2009-2017
Source: Homicide rates from Police of Guatemala. Migration, from EMIF Sur, weighted data. Population and poverty rates from National Institute of Statistics in Guatemala.

## 5. Methodology

### 5.1 Empirical Strategy

In this section we evaluate the effect that violence has on emigration flows. To do so, we estimate the total number emigration crossings made by Guatemalan residents at municipality level in Guatemala on an annual basis during 2009-2017. We use the sampling weights provided by the EMIF Sur, so these crossings are representative of the emigration flows in the border. We use the panel fixed effect specification shown in eq. (1). Since the majority of the emigration flows are intended to Mexico, we focus on these flows.

$$crossings_{mt}= \gamma_1+ \gamma_2 Poverty_{mt} + \gamma_3 DifhomicidesGuatemalaMexico_{mt}+ \gamma_4 year_t+ \gamma_5 mun_m +e_{mt} \quad (1)$$

where the dependent variable measures the number of emigration crossings that originated from municipality $m$ in Guatemala in year $t$. We separately analyse the number of emigration crossings made by Guatemalan residents to Mexico by adults, children and unaccompanied children. The regression coefficient $\gamma_2$ refers to the association with the poverty rate (at department level in Guatemala). $\gamma_3$ represents our main covariate, the difference between the homicides rate at municipality level in Guatemala and the national homicide rate in Mexico. We include time and municipality fixed-effects, *year*, *mun,* and use robust standard errors clustered at the municipality level.

Our panel fixed-effects regression specification helps to mitigate some concerns with endogeneity, particularly for any omitted time-invariant regressors that might be correlated with the error term. Nonetheless, we acknowledge that a limitation of this approach could be a potential reverse causality between migration and the differential in homicide rates between Guatemalan municipalities and Mexico. To test and address for such endogeneity bias we also use a panel fixed-effects model with two-stage least squares (IV-2SLS) instrumental variables. Eq. (2) represents the first-stage regressions of our potential endogenous variable, $\mu_3 DifhomicidesGuatemalaMexico$.

$$DifhomicidesGuatemalaMexico_{mt}=\kappa_1+ \kappa_2 Z_{mt}+ \kappa_3 Poverty_{mt}+ \kappa_4 year_t+ \kappa_5 mun_m+v_{mt} \quad (2)$$

where $v_{mt}$ refer to the disturbance term. We use two instruments, denoted by $Z_{mt}$. The first instrument is the kilograms seized of cocaine paste in Colombia, on annual basis, during our period of analysis. We have this information on annual basis, and as instrument we use the lagged value of the cocaine paste seized. As second instrument, we use the interaction between the amount of seized cocaine paste in Colombia (lagged by one period) and the annual number of hectares deforested in Guatemala at municipality level.

Cocaine paste is an intermediary product in the chemical extraction of cocaine from coca leaves, thus its seizure in Colombia will affects the overall supply of cocaine in the global market. Roughly 86% of the cocaine trafficked globally moves through Central America via drug trafficker organizations that operate in at least 50% of Guatemalan territory (UNODC 2010). Our rationale is that shocks in the supply of cocaine, driven by the seizures of cocaine paste in Colombia, will make the cocaine market more valuable and increase the chances that drug cartels will fight to control trafficking routes in Guatemala. Since changes in the supply of cocaine might have a delayed effect on violence, we use this variable lagged by one year. We interact this variable with deforestation because drug trafficking accounts for 15-30% of annual national deforestation in Central America (Nellemann 2012; Sesnie et al. 2017). Such deforestation practices can fuel even more violence due to loss of livelihoods and land. As shown in Figure 6, the interaction between the lagged seizures of paste cocaine and deforestation is positively correlated with the differential in local homicide rates in Guatemala and the average homicide rates in Mexico.[4]

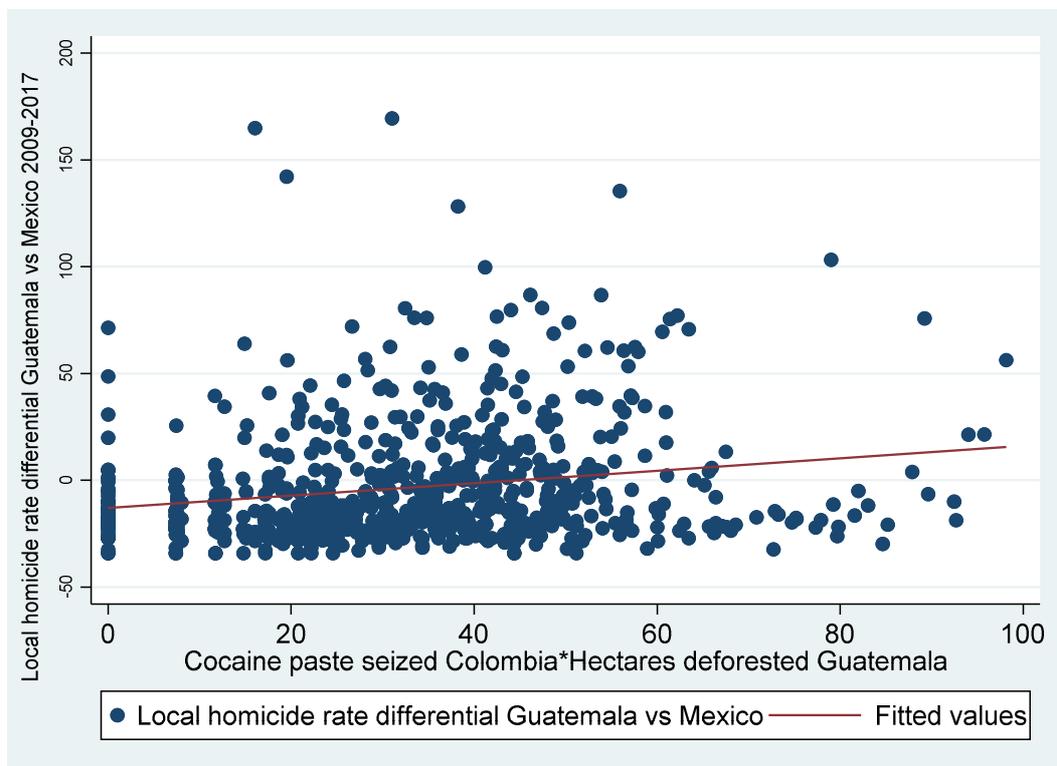

Fig 6. Local homicide rate differential between Guatemala and Mexico and deforestation at municipality level in Guatemala during 2009-2017

---

[4] We obtained the seizures of cocaine paste from the Colombian Observatory of Drugs. The number of deforested units at municipality level, on an annual basis, was taken from the University of Maryland Department of Geographical Sciences Global Change (Hansen et al. 2013).

The second-stage IV fixed-effects model estimates the impact on the differential in homicide rates between Guatemala and Mexico on emigration flows, as shown in eq. (3).

$$crossings_{mt} = \varphi_1 + \varphi_2 Poverty_{mt} + \varphi_3 \widehat{DifhomicidesGuatemalaMexico}_{mt} + \varphi_4 year_t + \varphi_5 mun_{mt} + \varphi_m + \varepsilon_{mt}$$

(3)

where $\varphi_3$ is the regression coefficients of the instrumented endogenous variable. We include as additional controls poverty rate, year and municipality fixed-effects. The terms $\varphi_m$ and $\varepsilon_{mt}$ represent the time-varying and -invariant residuals. We use robust standard errors clustered at the municipality level.

## 6. Results

Table 3 shows the panel fixed-effects results. There is no evidence that rises in local homicide rates at municipality level in Guatemala, relative to homicide rates in Mexico, are associated with emigration crossings. That is the case across all specifications ran including for adults, men, women, children, unaccompanied children. As mentioned earlier, a concern with these results is the potential endogeneity between emigration crossings and the differential in local violence in Guatemala and Mexico. If large flows of population leave Guatemala each year, this could have an impact on local violence. For instance, violence could decline if substantial flows of vulnerable young men and children emigrate, reducing the number of recruits available for local gangs. It is also possible that violence could increase as emigration flows rise. Thus, to test and correct, if necessary, for endogeneity we use the second-stage instrumental specifications mentioned in eq.(2)-eq.(3).

Table 3. Emigration flows during 2009-2017, panel fixed-effects at municipality level

|  | (1) | (2) | (3) | (4) | (5) | (6) | (7) | (8) | (9) |
|---|---|---|---|---|---|---|---|---|---|
|  | Emigrating from Guatemala to Mexico | | | | | | Emigrating from Guatemala to Mexico or the USA | | |
|  | Adults | Men | Children | Unaccompanied children | Women | Girls | Unaccompanied children | Unaccompanied boys | Unaccompanied girls |
| Difference homicide rate in Guatemalan municipality and average in Mexico | -11.960 | -24.863 | 2.029 | 1.138 | 12.903 | 0.088 | 1.069 | 1.059 | 0.010 |
|  | (37.124) | (34.784) | (1.441) | (0.856) | (10.794) | (0.148) | (0.850) | (0.944) | (0.125) |
| Log Poverty in Guatemala (Department) | 4,046.725 | 932.282 | 76.037 | 62.399 | 3,114.443 | 14.473 | 48.395 | 23.298 | 25.097 |
|  | (4,909.298) | (2,531.298) | (214.776) | (188.226) | (2,756.318) | (45.130) | (201.233) | (177.676) | (37.403) |
| Constant | -12,181.526 | -48.701 | -176.539 | -157.330 | -12,132.825 | -39.586 | -95.257 | -10.146 | -85.111 |
|  | (20,304.202) | (10,468.204) | (886.787) | (777.819) | (11,406.902) | (187.713) | (832.866) | (734.043) | (156.442) |
| Observations | 1,161 | 1,161 | 1,161 | 1,161 | 1,161 | 1,161 | 1,161 | 1,161 | 1,161 |
| R-squared | 0.007 | 0.018 | 0.043 | 0.035 | 0.017 | 0.009 | 0.033 | 0.039 | 0.012 |
| Number of municipalities | 251 | 251 | 251 | 251 | 251 | 251 | 251 | 251 | 251 |
| Municipality fixed effects | Yes | Yes | Yes | Yes | Yes | Yes | Yes | Yes | Yes |
| Year fixed effects | Yes | Yes | Yes | Yes | Yes | Yes | Yes | Yes | Yes |
| rho | 0.747 | 0.737 | 0.667 | 0.642 | 0.699 | 0.575 | 0.649 | 0.589 | 0.625 |
| Overall R-square | 0.00202 | 5.82e-05 | 0.00360 | 0.00351 | 0.000967 | 0.00177 | 0.00323 | 0.00423 | 0.00377 |

Note: Robust standard errors clustered at municipality level in parentheses. Significance level *** p<0.01, ** p<0.05, * p<0.1. Source: EMIF Sur, weighted data. Coverage: 2009-2017.

*6.1 Assessing the Impact of Differential in Homicide Rates on Emigration*

The first-stage regressions in Table A.1 column 1 in the Appendix show that the two instruments are strongly associated with the endogenous regressor. That is, more seizing of cocaine paste in Colombia, leads a year after to higher differences in the local violence in Guatemala, relative to the national average in Mexico. More deforestation in Guatemala, also widens the differences in local violence in Guatemala, relative in Mexico, particularly with increased seizures of cocaine paste in Colombia. The F-of these excluded instruments is above 10.

Table 4 presents the second-stage IV regressions. The bottom rows show the endogeneity tests and the Sargan-Hansen over-identification tests. These tests suggest the instruments used are valid. That is uncorrelated with the error term, and correctly excluded from the estimated equation. We also find evidence of endogeneity across all models, columns (1)-(4). These results refer to the number of emigration crossings made by adults, male adults, children, and unaccompanied children. These endogeneity tests suggest therefore that the IV panel fixed-effects regressions should be preferred to the panel fixed-effects model.

The panel fixed IV models suggest that a one-point increase in the differential homicides rate in Guatemalan municipalities, relative to the average rate in Mexico, leads to 246 additional emigration crossings made by adults, and specifically 211 made by males. Most of these emigration crossings were undocumented during our period of analysis, at least 60%. Column (3) suggests that a one-point increase in local homicide rates, compared to the average in Mexico leads to 20 additional crossings made by children, and from those 13 would be made by unaccompanied children.

*6.2 Results for Women and Girls*

It is crucial to include gender-based violence approaches in the discussions about emigration and displacement. As mentioned earlier only 10% of adult emigrants are female. There are many reasons for such low percentage of female emigrants travelling from Guatemala. Females often have more caring responsibilities for their younger siblings, parents and children, which limits their ability to travel. Females also face much greater risks of violence and sexual assault during their journey north, which can deter them. Females are also less likely to find employment in Mexico, particularly in the agricultural sector, which would not allow them to support their families back home. For similar reasons, only 10% of emigrant children are girls.

Table 4. Emigration flows of adults and children. Second-stage IV panel fixed-effects at municipality level

|  | (1) | (2) | (3) | (4) |
|---|---|---|---|---|
|  | Emigrating from Guatemala to Mexico | | | |
|  | Adults | Men | Children | Unaccompanied children |
| Difference homicide rate in Guatemalan municipality and average in Mexico | 246.351*** | 211.239*** | 20.250*** | 13.264*** |
|  | (86.352) | (66.090) | (5.053) | (3.515) |
| Log Poverty in Guatemala (department level) | 8,948.714 | 3,840.635 | 113.720 | 88.155 |
|  | (7,509.463) | (4,091.474) | (453.351) | (381.115) |
| Constant | -30,406.630 | -10,007.575 | -196.072 | -171.996 |
|  | (31,281.328) | (17,043.449) | (1,880.951) | (1,584.712) |
| Observations | 728 | 728 | 728 | 728 |
| Number of municipalities | 149 | 149 | 149 | 149 |
| Municipality fixed effects | Yes | Yes | Yes | Yes |
| Year fixed effects | Yes | Yes | Yes | Yes |
| Sargan-Hansen statistic | 1.264 | 2.533 | 2.462 | 2.073 |
|  P-value | 0.261 | 0.112 | 0.117 | 0.150 |
| Stock-Yogo weak ID test critical values: 15% maximal IV size | 12.701 | 12.701 | 12.701 | 12.701 |
| Weak identification test (Cragg-Donald Wald F statistic) | 11.590 | 11.590 | 11.590 | 11.590 |
| Davidson-MacKinnon test of exogeneity | 7.571 | 10.045 | 9.168 | 6.068 |
|  P-value | 0.006 | 0.002 | 0.003 | 0.014 |

Note: Homicide rate and difference in homicide rates instrumented with the lagged kilograms seized of cocaine paste in Colombia, and its interaction with deforested hectares in Guatemala. Robust standard errors clustered at municipality level in parentheses. First-stage instrumental variable model in Table A.1, column 1. Significance level *** $p<0.01$, ** $p<0.05$, * $p<0.1$.

Source: EMIF Sur, weighted data. Coverage: 2009-2017.

Table 5. Number of emigration crossings made by women, unaccompanied boys and girls. Second-stage IV panel fixed-effects at municipality level

|  | (1) | (2) | (3) | (4) | (5) |
|---|---|---|---|---|---|
|  | Emigrating from Guatemala to Mexico | | Emigrating from Guatemala to Mexico or the USA | | |
|  | | | Unaccompanied children | Unaccompanied boys | Unaccompanied girls |
|  | Women | Girls | | | |
| Difference homicide rate in Guatemalan municipality and average in Mexico | 35.112 | -0.253 | 12.247*** | 13.155*** | -0.908 |
|  | (26.318) | (0.702) | (3.574) | (3.646) | (0.689) |
| Log Poverty in Guatemala (department level) | 5,108.079 | 18.809 | 47.311 | 17.232 | 30.079 |
|  | (4,237.436) | (83.890) | (402.557) | (369.639) | (76.095) |
| Constant | -20,399.055 | -20,399.06 | 1.225 | 103.758 | -102.533 |
|  | (17,679.226) | -17,679.23 | (1,674.851) | (1,534.907) | (317.773) |
| Observations | 728 | 728 | 728 | 728 | 728 |
| Number of municipalities | 149 | 149 | 149 | 149 | 149 |
| Municipality fixed effects | Yes | Yes | Yes | Yes | Yes |
| Year fixed effects | Yes | Yes | Yes | Yes | Yes |
| Sargan-Hansen statistic | 0.379 | 0.308 | 2.175 | 2.774 | 0.304 |
| P-value | 0.538 | 0.579 | 0.141 | 0.096 | 0.581 |
| Stock-Yogo weak ID test critical values: 15% maximal IV size | 12.701 | 12.701 | 12.701 | 12.701 | 12.701 |
| Weak identification test (Cragg-Donald Wald F statistic) | 11.590 | 11.590 | 11.590 | 11.590 | 11.590 |
| Davidson-MacKinnon test of exogeneity | 0.358 | 0.047 | 5.146 | 7.830 | 0.431 |
| P-value | 0.550 | 0.828 | 0.024 | 0.005 | 0.512 |

Note: Homicide rate and difference in homicide rates instrumented with the lagged kilograms seized of cocaine paste in Colombia, and its interaction with deforested hectares in Guatemala. Robust standard errors clustered at municipality level in parentheses. First-stage instrumental variable model in Table A.1, column 1. Significance level *** p<0.01, ** p<0.05, * p<0.1.

Source: EMIF Sur, weighted data. Coverage: 2009-2017.

We study next whether emigration crossings made by women and girls are also impacted by rises in local homicides in Guatemala, relative to Mexico. We re-run the panel fixed-effects IV regressions, using the same instruments as before. The first-stage IV regression is the same as the one shown in Table A.1, column 1. The second-stage IV results, in Table 5, show that the instruments are valid. However, we fail to find evidence of endogeneity for the emigration crossings made by women and girls travelling to Mexico (columns 1 and 2). We also fail to find any statistically significant evidence that rises in local homicide rates in Guatemala, relative to the average rate in Mexico, affect the number of crossings made by women or girls.

We also analyze the number of crossings made by unaccompanied children made to Mexico or the USA. There is evidence of endogeneity only for the whole group of unaccompanied children, unaccompanied boys, but not unaccompanied girls (Table 5, columns 3, 4 and 5). The IV panel fixed-effects results suggest that a one one-point increase in the differential homicides rate in Guatemalan municipalities, relative to the average rate in Mexico, leads to 12.247 additional emigration crossings made by unaccompanied children going to Mexico or the USA, and 13 made by unaccompanied boys. There is no impact on the crossings made by unaccompanied girls.

Our findings thus far suggests that rises in local homicide rates, relative to average in Mexico, does not lead to more emigration from girls and women. There are many economic and socio-economic reasons why more men and boys choose to emigrate illegally. Although women and girls might also have these economic motivations to migrate, they face higher costs and risks for emigrating than men. Also, we acknowledge that there are many other ways in which violence could affect women and girls not fully captured by changes in homicide rates. They may be subjected to crime, violence and sexual exploitation by gangs or smugglers. We explore this issue further in the next section.

## 7. Robustness checks

### 7.1 Alternative Instruments: Retail Price of Cocaine

Our identification strategy has argued that seizing cocaine paste in Colombia leads to supply shocks of cocaine, and indirectly increases cocaine prices, profitability of trafficking drugs in Central America and violence in the region. As a robustness check, we slightly change our instruments. We use the same interaction between deforestation in Guatemalan municipalities and kilograms of cocaine paste seized in Colombia. We also add the retailing price of cocaine per gram in USA, adjusted for purity and inflation in dollars. The first-stage IV regression is shown in Table A.1, column 2. Both our instruments are positive and statistically significant. The F-statistic of these excluded instruments is above 10.

Table 6 shows the second stage IV panel fixed-effects for adults, men, children and unaccompanied children traveling from Guatemala to Mexico. The Sargan-Hansen statistics suggest the instruments are valid. We find endogeneity for all the models, with the exception of all adults. Our IV findings suggest that a one-point increase in the differential homicides rate in Guatemalan municipalities, relative to the average rate in Mexico, leads to almost 200 additional emigration crossings made by male adults, 20 children and 13 unaccompanied children. All these findings are very similar in magnitude to our earlier findings, albeit we acknowledge the instruments are not as strong as our earlier specification as the F-statistic is slightly smaller a, the Stock-Yogo suggest we might have also a slightly higher bias.

Table 6. Emigration flows of adults and children. Using change in price of retail of cocaine as alternative instruments, second-stage IV panel fixed-effects at municipality level

|  | (1) | (2) | (3) | (4) |
|---|---|---|---|---|
|  | Emigrating from Guatemala to Mexico | | | |
|  | Adults | Men | Children | Unaccompanied children |
| Difference homicide rate in Guatemalan municipality and average in Mexico | 233.079*** | 198.771*** | 20.425*** | 13.440*** |
|  | (85.855) | (65.886) | (5.204) | (3.666) |
| Log Poverty in Guatemala (department level) | 8,807.649 | 3,708.122 | 115.584 | 90.026 |
|  | (7,491.144) | (4,052.063) | (455.567) | (383.349) |
| Constant | -29,867.906 | -9,501.510 | -203.193 | -179.142 |
|  | (31,206.788) | (16,879.462) | (1,889.736) | (1,593.496) |
| Observations | 728 | 728 | 728 | 728 |
| Number of municipalities | 149 | 149 | 149 | 149 |
| Municipality fixed effects | Yes | Yes | Yes | Yes |
| Year fixed effects | Yes | Yes | Yes | Yes |
| Sargan-Hansen statistic | 1.28 | 2.62 | 2.46 | 2.08 |
| P-value | 0.26 | 0.11 | 0.12 | 0.15 |
| Stock-Yogo weak ID test critical values: 15% maximal IV size | 11.65 | 11.65 | 11.65 | 11.65 |
| Weak identification test (Cragg-Donald Wald F statistic) | 11.59 | 11.59 | 11.59 | 11.59 |
| Davidson-MacKinnon test of exogeneity | 6.29 | 8.31 | 8.59 | 5.74 |
| P-value | 0.12 | 0.00 | 0.00 | 0.02 |

Note: Homicide rate and difference in homicide rates instrumented with the retail price of price of cocaine to consumers in the USA, and the interaction to lagged kilograms seized of cocaine paste in Colombia, and deforested hectares in Guatemala. Robust standard errors clustered at municipality level in parentheses. First-stage instrumental variable model in Table A.1, column 2. Significance level *** p<0.01, ** p<0.05, * p<0.1. Source: EMIF Sur, weighted data. Coverage: 2009-2017.

We also re-run our results for women and girls travelling to Mexico. The first-stage IV regression is also presented in Table A.1, column 2. The second-stage IV panel fixed effects, suggest there is no evidence of endogeneity (Table A.2, columns 1 and 2). Once again the panel fixed-effects IV models suggest that increases in local homicides, relative to the average level in Mexico, do not affect emigration of girls nor women.

We also re-run our IV results for unaccompanied children emigrating from Mexico or the USA. As before, we find that rises in local homicide rates in Guatemala, relative to Mexico increases emigration crossings of unaccompanied children by nearly 12 crossings and by boys by 13. Once again, only for these two groups we find evidence of endogeneity (Table A.2, columns 3 and 4). For the case of girls, as our earlier specifications, we find no evidence of endogeneity and no evidence that rises in local homicide rates, relative to Mexico, affect their emigration crossings (column 5).

*7.2 Alternative Measure of Local Violence: Theft Rate in Guatemala*

It is unclear why women and girls in Guatemala seem unaffected by increases in local homicide rates, relative to the average rate in Mexico, while men and boys are. This contrasting impacts perhaps could be driven by males being more likely to being killed by homicide than women. However, we acknowledge that women experience other forms of violence that might also affect their decision to emigrate. Thus, as a robustness test, we analyze whether increases in criminality in Guatemala leads to higher emigration crossings.

Crime statistics are imperfect, particularly in settings with high impunity where people have little incentives to even report incidents. For this reason, our analysis thus far has relied on homicide rates as a more vigorous measure of violence. However, as a robustness check we use the local rate of theft in Guatemala as a proxy of which areas experience from more criminality. This statistic is available only at department level. Still, we maintain our emigration figures aggregated at the lower municipality level.

Since the impact of local theft rate on emigration can be endogenous, we use panel fixed-effects IV variables. We use the very first set of instruments as they are more strongly related to theft rate. These instruments are the kilos of seized cocaine paste in Colombia (lagged by one year), and its interaction with local deforestation in Guatemala. The first-stage IV results are shown in Table A.1, column 3. The two instruments are statistically significant, and positively correlated with theft rate, as expected. The F-statistic of the excluded instruments is nearly 39.

The second-stage IV panel fixed effects results are shown in Table A.3. We find that increases in the theft rate have no impact on the emigration crossings made by women and girls when analyzing these groups separately (columns 5 and 6). In contrast, increases in local theft rates in Guatemala leads to more emigration crossings made by men, children, and unaccompanied children from Guatemala to Mexico (columns 1-4). The magnitude of the coefficients is smaller than our earlier results using homicide rates. These differences could stem from theft rate being measured at a more aggregate level, department, and also being subject to more underreporting of this type of crime than homicides. Also worth noting, is that higher theft rate increases the number of crossings of unaccompanied children and boys to Mexico or the USA, but not of unaccompanied girls (Table A.3, columns 7-9).

## 8. Conclusions

In this paper, we analyzed the factors driving emigration from Guatemala to Mexico. Our analysis offered four key findings. First, according to the EMIF Sur, the largest surveys of emigration in the region the majority of emigration crossings are made by migrants who claim they are traveling to Mexico for work purposes for longer than a month, and as their final intended destination. These migrants seem to be seasonal border workers as at the moment of the interview they had already made an average of 70 crossings to Mexico for work purposes The frequent number of trips can also be explained as most migrants live between the border of Guatemala and Mexico. Despite the proximity, most of them (at least 60%) travel to Mexico undocumented.

Second, we found that increases in the homicide rates in Guatemala relative to Mexico, have a substantial impact on adult emigration. A one-point increase in the differential homicides rate in Guatemalan municipalities, relative to the average rate in Mexico, leads to 211 additional emigration crossings made by male adults. This is an important finding as much of the literature has focused on child migration (Amuedo-Dorantes et al., 2015; Clemens, 2021).

Third, amongst all the children emigrating, 99% left seeking jobs in Mexico. Most of them (70%) left unaccompanied. These children were pushed by increases in local homicide rates in Guatemala, relative to the average levels in Mexico. We found that a one-point increase in the differential homicides rate in Guatemalan municipalities, relative to the average rate in Mexico, leads to 20 additional crossings made by children, and from those 13 would be made by unaccompanied children.

Fourth, as a robustness check we also estimated the impact of increases in violence in Guatemala using alternative instruments and proxies. Once again, our findings suggested that rises in local homicide rates, relative to the average level in Mexico leads to more emigration crossings of men, children and unaccompanied children. We found no impact on women or girls. These results remain robust to using changes in the local theft rate in Guatemala.

Earlier literature has shown that rises in local violence in Guatemala is associated to more Central American children being apprehended at the USA border (Clemens 2021). Our findings add to this literature that not all child migrants from Central America travel to the USA. A substantial number become seasonal border workers in Mexico. Evidently, although not disclosed, some of these migrants might attempt to cross the USA border subsequently as substantiated by the large number of deportations at this border. Still, the bulk of these migrants intend to stay in Mexico for prolonged periods and for work reasons (Gutiérrez-Romero and Salgado 2022).

The USA-Mexican enforcement-centric approach has not deterred illegal migration and not dealt with the bottom of the migration and security problem in the region (Amuedo-Dorantes, Pozo, and Puttitanun 2015; Clemens 2021). Drug-trafficking violence and environmental conflicts continue driving emigration flows. Thus, more comprehensive immigration policies are needed in the region, tackling the underlying roots of deprivation and violence.

# Appendix

Table A.1 First-stage IV regressions of Tables 4, 5, 6, A.2 and A.3

|  | (1) Model 1 Difference homicide rate in Guatemalan municipality and average in Mexico | (2) Model 2 Difference homicide rate in Guatemalan municipality and average in Mexico | (3) Model 3 Theft rate at department level in Guatemala |
|---|---|---|---|
| Log hectareas deforested in Guatemalan municipalities x Log seized cocaine paste in Colombia lagged by one year | 0.154* | 0.152* | 0.179* |
|  | (0.090) | (0.090) | (0.092) |
| Log seized cocaine paste in Colombia, lagged by one year | 127.810*** |  | 192.048*** |
|  | (22.601) |  | (23.993) |
| Cocaine street price per gram in USA, adjusted for purity and inflation in dollars |  | 0.853*** |  |
|  |  | (0.159) |  |
| Log poverty in Guatemala (department level) | -12.065 | -11.915 | -38.809 |
|  | (10.811) | (10.859) | (27.838) |
| Constant | -1,325.524*** | -43.702 | -1,836.091*** |
|  | (244.101) | (47.361) | (274.206) |
| Observations | 728 | 728 | 728 |
| R-squared | 0.075 | 0.072 | 0.246 |
| Number of municipalities | 149 | 149 | 149 |
| Municipality fixed effects | Yes | Yes | Yes |
| Year fixed effects | Yes | Yes | Yes |
| F-statistic of excluded instruments | 11.06 | 10.05 | 38.86 |
| p-value | 0.000 | 0.000 | 0.000 |

Note: Robust standard errors clustered at municipality level in parentheses. Second-stage instrumental variables in Tables 4-8. Significance level *** p<0.01, ** p<0.05, * p<0.1. Source: EMIF Sur, weighted data. Coverage: 2009-2017.

Table A.2. Emigration flows of women, girls and unaccompanied children. Using change in price of retail of cocaine as alternative instruments, second-stage IV panel fixed-effects at municipality level

| | (1) | (2) | (3) | (4) | (5) |
|---|---|---|---|---|---|
| | \multicolumn{2}{c}{Emigrating from Guatemala to Mexico} | | \multicolumn{3}{c}{Emigrating from Guatemala to Mexico or the USA} | | |
| | Women | Girls | Unaccompanied children | Unaccompanied boys | Unaccompanied girls |
| Difference homicide rate in Guatemalan municipality and average in Mexico | 34.308 | -0.318 | 12.440*** | 13.414*** | -0.974 |
| | (25.821) | (0.779) | (3.757) | (3.802) | (0.767) |
| Log Poverty in Guatemala (department level) | 5,099.527 | 18.115 | 49.370 | 19.985 | 29.385 |
| | (4,232.262) | (84.628) | (404.788) | (371.858) | (76.940) |
| Constant | -20,366.396 | -50.806 | -6.637 | 93.247 | -99.884 |
| | (17,659.772) | (353.064) | (1,683.619) | (1,543.775) | (321.032) |
| Observations | 728 | 728 | 728 | 728 | 728 |
| Number of municipalities | 149 | 149 | 149 | 149 | 149 |
| Municipality fixed effects | Yes | Yes | Yes | Yes | Yes |
| Year fixed effects | Yes | Yes | Yes | Yes | Yes |
| Sargan-Hansen statistic | 0.38 | 0.31 | 2.19 | 2.80 | 0.31 |
| P-value | 0.54 | 0.58 | 0.14 | 0.09 | 0.58 |
| Stock-Yogo weak ID test critical values: 15% maximal IV size | 11.65 | 11.65 | 11.65 | 11.65 | 11.65 |
| Weak identification test (Cragg-Donald Wald F statistic) | 11.59 | 11.59 | 11.59 | 11.59 | 11.59 |
| Davidson-MacKinnon test of exogeneity | 0.31 | 0.06 | 4.90 | 7.53 | 0.46 |
| P-value | 0.58 | 0.80 | 0.03 | 0.01 | 0.50 |

Note: Homicide rate and difference in homicide rates instrumented with the retail price of price of cocaine to consumers in the USA, and the interaction to lagged kilograms seized of cocaine paste in Colombia, and deforested hectares in Guatemala. Robust standard errors clustered at municipality level in parentheses. First-stage instrumental variable model in Table A.1, column 2. Significance level *** p<0.01, ** p<0.05, * p<0.1. Source: EMIF Sur, weighted data. Coverage: 2009-2017.

Table A.3. Impact of rate of theft on emigration flows of women, girls and unaccompanied children. Second-stage IV panel fixed-effects at municipality level

| | (1) | (2) | (3) | (4) | (5) | (6) | (7) | (8) | (9) |
|---|---|---|---|---|---|---|---|---|---|
| | \multicolumn{6}{c}{Emigrating from Guatemala to Mexico} | | \multicolumn{3}{c}{Emigrating from Guatemala to Mexico or the USA} | | |
| | Adults | Men | Children | Unaccompanied children | Women | Girls | Unaccompanied children | Unaccompanied boys | Unaccompanied girls |
| Theft rate in Guatemalan departments | 172.583*** | 150.436*** | 14.151*** | 9.305*** | 22.147 | -0.212 | 8.619*** | 9.264*** | -0.645 |
| | (58.965) | (48.406) | (3.922) | (2.848) | (15.081) | (0.488) | (2.881) | (2.926) | (0.490) |
| Log Poverty in Guatemala (department level) | 12,702.949 | 7,150.275 | 421.020 | 290.741 | 5,552.674 | 13.676 | 235.409 | 219.483 | 15.926 |
| | (8,698.151) | (5,400.995) | (589.205) | (460.905) | (4,469.346) | (90.216) | (478.372) | (456.897) | (87.321) |
| Constant | -57,698.715 | -33,939.287 | -2,431.882 | -1,644.100 | -23,759.428 | -17.960 | -1,364.109 | -1,364.026 | -0.082 |
| | (39,044.294) | (24,128.196) | (2,457.940) | (1,936.013) | (19,635.883) | (396.058) | (2,031.492) | (1,924.008) | (390.637) |
| Observations | 728 | 728 | 728 | 728 | 728 | 728 | 728 | 728 | 728 |
| Number of municipalities | 149 | 149 | 149 | 149 | 149 | 149 | 149 | 149 | 149 |
| Municipality fixed effects | Yes | Yes | Yes | Yes | Yes | Yes | Yes | Yes | Yes |
| Year fixed effects | Yes | Yes | Yes | Yes | Yes | Yes | Yes | Yes | Yes |
| Sargan-Hansen statistic | 1.20 | 3.13 | 1.31 | 1.01 | 0.45 | 0.29 | 1.09 | 1.34 | 0.24 |
| P-value | 0.27 | 0.08 | 0.25 | 0.32 | 0.50 | 0.59 | 0.30 | 0.25 | 0.62 |
| Stock-Yogo weak ID test critical values: 15% maximal IV size | 33.12 | 33.12 | 33.12 | 33.12 | 33.12 | 33.12 | 33.12 | 33.12 | 33.12 |
| Weak identification test (Cragg-Donald Wald F statistic) | 11.59 | 11.59 | 11.59 | 11.59 | 11.59 | 11.59 | 11.59 | 11.59 | 11.59 |
| Davidson-MacKinnon test of exogeneity | 4.08 | 5.60 | 13.41 | 8.48 | 0.11 | 0.16 | 7.36 | 11.44 | 0.82 |
| P-value | 0.04 | 0.18 | 0.00 | 0.00 | 0.74 | 0.69 | 0.01 | 0.00 | 0.36 |

Note: Theft rate in Guatemalan districts instrumented with the retail price of price of cocaine to consumers in the USA, and the interaction to lagged kilograms seized of cocaine paste in Colombia, and deforested hectares in Guatemala. Robust standard errors clustered at municipality level in parentheses. First-stage instrumental variable model in Table A.1, column 3. Significance level *** p<0.01, ** p<0.05, * p<0.1. Source: EMIF Sur, weighted data. Coverage: 2009-2017.